# A UNIVERSAL INTEGRAL SCALING OF THE MEAN VELOCITY PROFILES IN TURBULENT WALL-BOUNDED FLOWS

T.-W. Lee

Department of Mechanical and Aerospace Engineering

Arizona State University

Tempe, Arizona, USA

attwl@asu.edu

**ABSTRACT**

Scaling of the mean velocity profiles has been studied by many researchers, since it provides a template of a universal dynamical pattern across a range of Reynolds numbers. Various normalization schemes have been shown in the past, some with a good degree of accuracy. An alternative, universal scaling is presented, where an integrated velocity profile serves as a universal template for incompressible, adverse pressure-gradient, and compressible flows.

**Introduction**

Scaling of the mean velocity profile serves an important function of compressing the dynamical effects that distribute the momentum in space, and therefore has been investigated by various researchers (e.g. Barenblatt et al. 1997; Marusic et al., 2010; McKeon et al., 2004; Schultz and Flack, 2013; Wei and Knopp, 2023; Zagarola and Smits, 1998). Once a universal scaling or self-similarity rule is found, then the mean velocity structure for a specific Reynolds number can be de-compressed or reconstructed. Some of the well-known scaling rules are the law of the wall and outer scaling (Tennekes and Lumley, 1972). For wall-bounded flows, the friction velocity ($u_\tau=(\tau_w/\rho)^{1/2}$) and the wall coordinate ($yu_\tau/\nu$) are used. For example, an assumption of viscous-dominated flow close to the wall leads to $u^+=y^+$. Also, use of the mixing length model gives the log-law: $u^+=1/\kappa \ln(y^+)+C$ (Tennekes and Lumley, 1972). Beyond this region, outer scaling such as Zagarola-Smits (2022) works quite well for boundary-layer flows. Thus, the scaling analysis involves different "rules" for inner, overlap, and outer regions. If the flow conditions change to compressible or adverse pressure-gradient (APG), then new scaling parameters are introduced. For example, van-Driest (1951) transform accounts for the density variation in compressible flows. Its modified version by Trettle and Larsson (2016) converts both the velocity and wall-normal coordinates, with a good degree of collapse for compressible channel flows. The velocity defect scaling by Wei and Knopp (2023) also leads to a self-similar velocity profile in the outer regions of APG flows.

In all of the different types of flows, the inner region is dominated by the viscous effects, so that the inner coordinate $y^+$ summarizes the viscous effect there. In this work, we report on an alternate scaling rule that involves integration of the velocity profiles with respect to the inner coordinate, $y^+$, but applicable from the wall to the centerline or the free-stream boundary condition. Also, this "integral" scaling appears to work for incompressible to compressible, and APG flows with minor modifications. In addition to providing a complete scaling rule, the integrated velocity reveals interesting and unifying properties of wall-bounded turbulent flows. The physical origin of the integral scaling for the mean velocity is also briefly contemplated in this work.

**Integral Scaling**

In our previous work, we have shown that gradient scaling works quite well in collapsing the Reynolds stress components (Lee, 2021a; Lee, 2021b; Lee and Park, 2024). For the mean velocity, an inverse operation, integration with respect to $y^+$ (Eq. 1), leads to a near-universal profile. Eq. 1 is a cumulative momentum integral from the wall to $y^+$, and is denoted as $I(U^+)$ as a function of $y^+$.

$$I(U^+) = \int_0^{y^+} U(y^+)dy^+ \qquad (1)$$

Figure 1 shows the integrated $U^+$ for incompressible channel flows for $Re_\tau$ =110 to 5200, using the DNS data by Iwamoto et al. (2002) and Lee and Moser (2015). The collapse of the velocity profiles at all Reynolds numbers is evident, from wall to the centerline. The $y^+$ coordinate is extended outward as the Reynolds number increases. There is a complete self-similarity in the integrated velocity profiles for incompressible channel flows. In this way, the wall coordinate is useful throughout the boundary layer, as the integration variable. There are two regions ($y^+ < 20$ and $y^+ > 20$): one with a slightly larger slope near the wall, and the outer region.

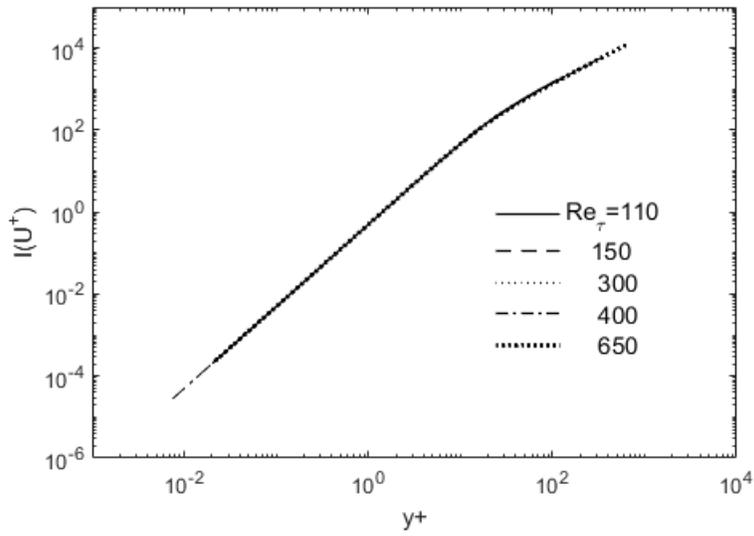

(a)

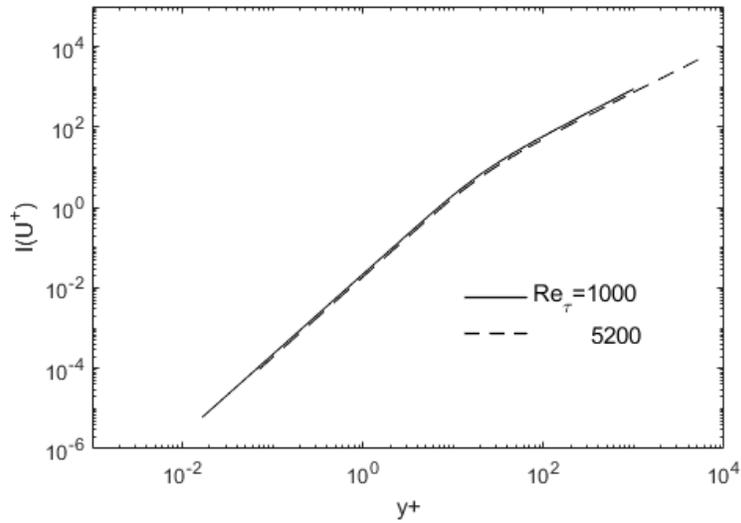

(b)

**Figure 1. Integrated U⁺ scaling for incompressible channel flows. The DNS data from Iwamoto et al. (2002), Re$_\tau$=110-650 (a), and Lee and Moser (2015), Re$_\tau$=1000 and 5200 (b), are used.**

For compressible flows, density variations introduce a complicating effect (Gerolymos and Vallet, 2022). The DNS data by Gerolymos and Vallet (2022) includes a variation of the centerline streamwise Mach number ($M_{CL}$) and friction Reynolds number ($Re_\tau$), from which we take six representative cases ($M_{CL}$ =0.8 – 2.11; $Re_\tau$=168 - 1478). Density- and viscosity-adjusted wall-normal coordinates ($y^*$) are used in Gerolymos and Vallet (2022). In Figure 2(a), the mean velocity profiles are plotted as a function of $y^*$. Except near the wall, the mean velocity profiles start to disperse ($y^* > 10$). Dual normalization of the both the velocity and wall-normal coordinate leads to a good scaling (Trettle and Larsson, 2016) for compressible channel flows. However, a simple integral operation (Eq. 1) also collapses mean velocity profiles across the entire boundary layer, as shown in Figure 2(b). There are again two segments in the integrated velocity in the log-log plot. If we parameterize them, then we will have a general mean velocity scaling to recover the profiles at any Reynolds number, for the entire flow width of the flow. van-Driest and related transforms also work quite well for compressible *channel* flows (Trettle and Larsson, 2016). Therefore, we look at compressible *boundary-layer* flow next.

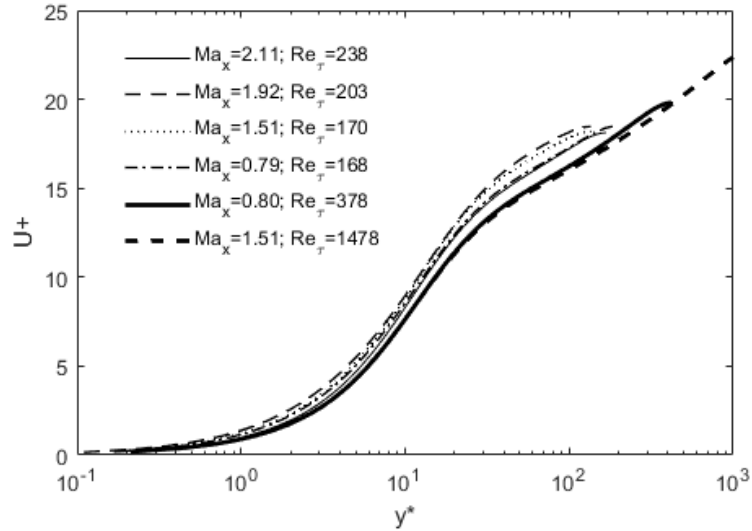

(a)

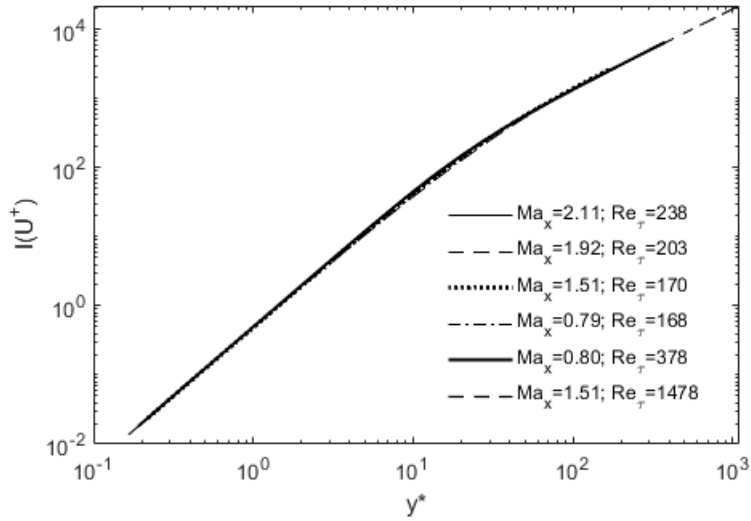

(b)

**Figure 2. Mean velocity (a) and integrated U⁺ (b) for compressible channel flows. The DNS data from Gerolymos and Vallet (2023) is used in Eq. 1.**

DNS data from Wenzel et al. (2018) cover Mach number from 0.3 to 2.5, and $Re_\tau$ from 250 to 450. We select the set at $Re_\tau$ = 359, and again apply the integral operation to $U^+$. A main difference is that the data are plotted as a function of $y^+$, in Wenzel et al. (2018). This alters the shape of the integrated velocity profiles, $I(U^+)$, but the self-similarity is retained with a curved shape on a log-log plot (Figure 3). The integration process appears to account for the density variation so that a universal profile is obtained.

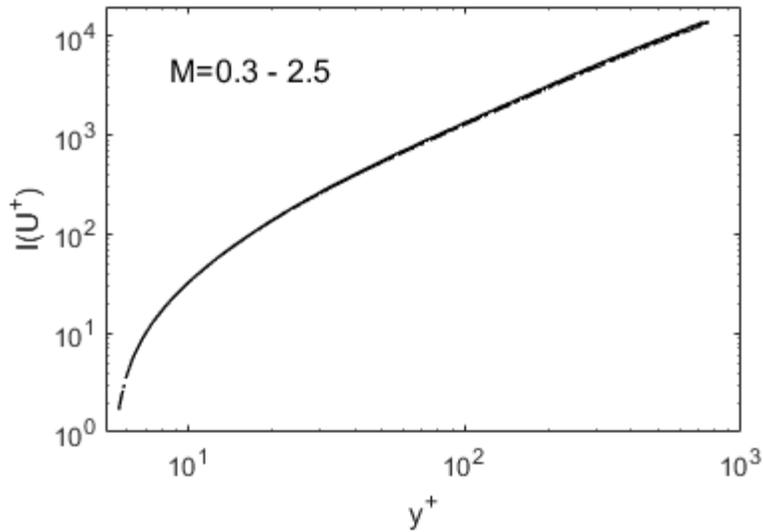

**Figure 3. Integrated $U^+$ for compressible boundary-layer flows. The DNS data from Wenzel et al. (2018). Mach number (M) ranges from 0.3 to 2.5, at $Re_\tau=359$.**

$U^+$ vs $y^+$ are plotted for zero (ZPG) and adverse pressure-gradient (APG) flows in Figure 4, using DNS data of Soria et al. (2017). ZPG and two APG conditions are computed in Soria et al. (2017): $\beta$ = 0, 1.9 and 39 (Soria et al., 2017). The adverse pressure-gradient reduces the velocity magnitude, while expanding the flow width. In order to account for the flow expansion, we re-scale $y^+$ with a boundary-layer width factor, $\delta^*$. The integrated velocity profiles using Eq. 1 and $y^+/\delta^*$ remain compact in Figure 5, until y+ ~ 20 where $\beta$ = 39 data start to deviate. At high $\beta$, there appears to be a departure from "equilibrium" momentum distribution that deviates from a common integrated velocity profile at large $y^+/\delta^*$.

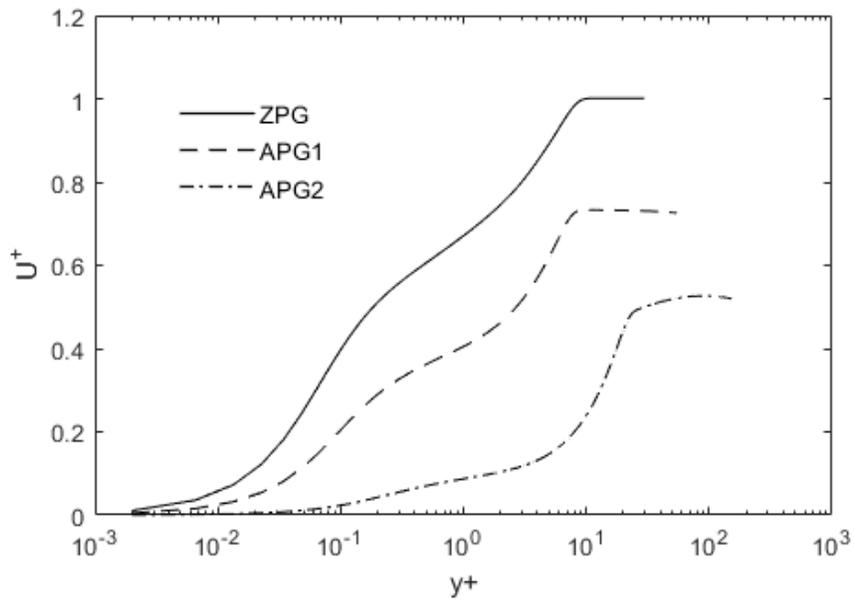

**Figure 4. Mean velocity (U+) in ZPG and APG flows. The DNS data from Soria et al. (2017) is used.**

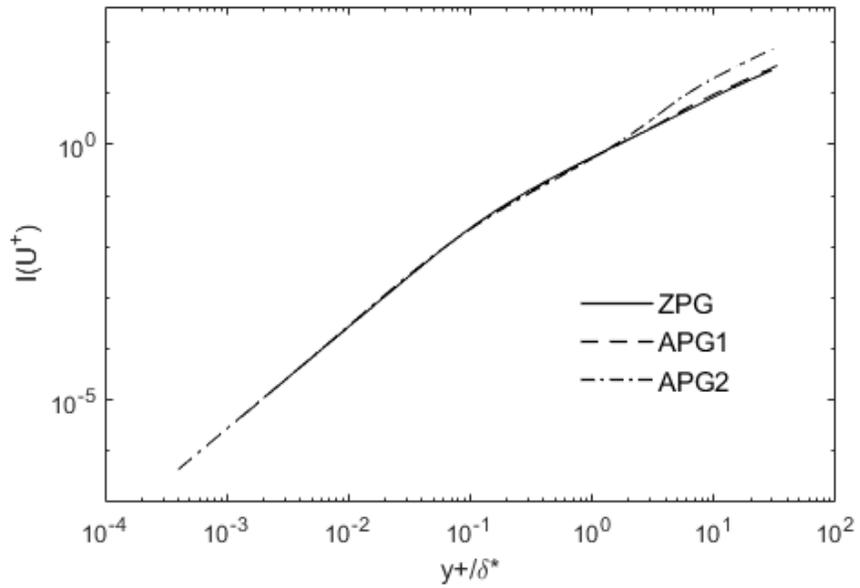

**Figure 5. Integrated U+ for ZPG and APG flows. The DNS data from Soria et al. (2017) is used.**

**Concluding Remarks**

An alternative mean velocity scaling involving an integral operation is shown to be universally valid for incompressible, adverse pressure-gradient and compressible wall-bounded flows. In Jimenez (2018), an energy integral of the mean velocity is considered, and the wall effects are shown to have far-reaching consequences throughout the boundary layer. We can also interpret the integrated velocity (Eq. 1) as the cumulative momentum content, normalized by the viscous effect. Therefore, this cumulative momentum integral remains universal at all Reynolds numbers, due to its accounting of the momentum relative to the viscous effect. The viscosity-adjusted momentum content is constant, and this volumetric or "regional" near-wall condition dictates the momentum distribution for the entire boundary layer. With the scaling parameter $y^+$ (or $y^*$ for compressible flows) the total mean momentum content (integrated $U^+$) relative to the viscosity is only redistributed in the cross-stream direction at different Reynolds numbers. From this universal integrated velocity profile, the mean velocity structure at any Reynolds numbers can be recovered simply by differentiating with respect to $y^+$ ($d/dy^+$) corresponding to that $Re_\tau$ ($u_\tau$ and $\nu$).